\documentclass[aip,apl,amsmath,amssymb,amsfonts,reprint,numerical]{revtex4-1}

\usepackage{graphicx}
\usepackage{dcolumn}
\usepackage{bm}
\usepackage{epsf}
\usepackage[usenames]{color}

\newcommand{\spderiv}[2]{\frac{\partial^2#1}{\partial#2^2}}      %
\newcommand{\pderiv}[2]{\frac{\partial#1}{\partial#2}}           %

\begin{document}

\title{Ultrasonic control of terahertz radiation via lattice anharmonicity in LiNbO$_{\bf 3}$}
\date{\today}

\author{R.~H. Poolman}
\author{E.~A. Muljarov}
\author{A.~L. Ivanov}
\affiliation{School of Physics and Astronomy, Cardiff University,
Cardiff CF24 3AA, United Kingdom}

\begin{abstract}
We propose a novel tunable terahertz (THz) filter using the
resonant acousto-optic (RAO) effect. We present a design based on
a transverse optical (TO) phonon mediated interaction between a
coherent acoustic wave and the THz field in LiNbO$_{3}$. We
predict a continuously tunable range of the filter up to 4\,THz
via the variation of the acoustic frequency between 0.1 and
1\,GHz. The RAO effect in this case is due to cubic and quartic
anharmonicities between TO phonons and the acoustic field. The
effect of the interference between the anharmonicities is also
discussed.
\end{abstract}

\maketitle

In recent years terahertz (THz) radiation has become a tool to
study a wide variety of phenomena, necessitating the development
of experimental techniques to access this spectral
band.\cite{Schmuttenmaer2004} Radiation in the THz range has been
used to study picosecond phonon dynamics, polariton propagation,
optical properties of various materials, and has been used for
imaging purposes.\cite{Feurer2003, Kojima2003, Han2001,
Stoyanov2002} There are also several commercial applications that
cover a variety of fields, including sub-mm wave astronomy,
chemical recognition and biomedical imaging for disease
diagnostics, THz imaging and sensing for security
applications.\cite{Siegel2002, Fischer2005, Federici2005} As a
result, spectrally resolved control of THz radiation has become an
important research topic. One implementation uses optical control
of carrier densities in type-I/type-II GaAs/AlAs multiple quantum
wells at cryogenic temperatures.\cite{Libon2000} Other
implementations include magnetically tuned liquid crystals in
metallic hole arrays and Lyot and Sloc filters, which were shown
to be tunable over various ranges between 0.1 and
$0.8$\,THz.\cite{Pan2005,Chen2006,Ho2008} A relative lateral
translation of two metallic photonic crystals has also resulted in
a THz filter (tunable between 0.365 and
0.385\,THz).\cite{Drysdale2004}
\begin{figure}[t]
    \includegraphics[angle=0,scale=0.3]{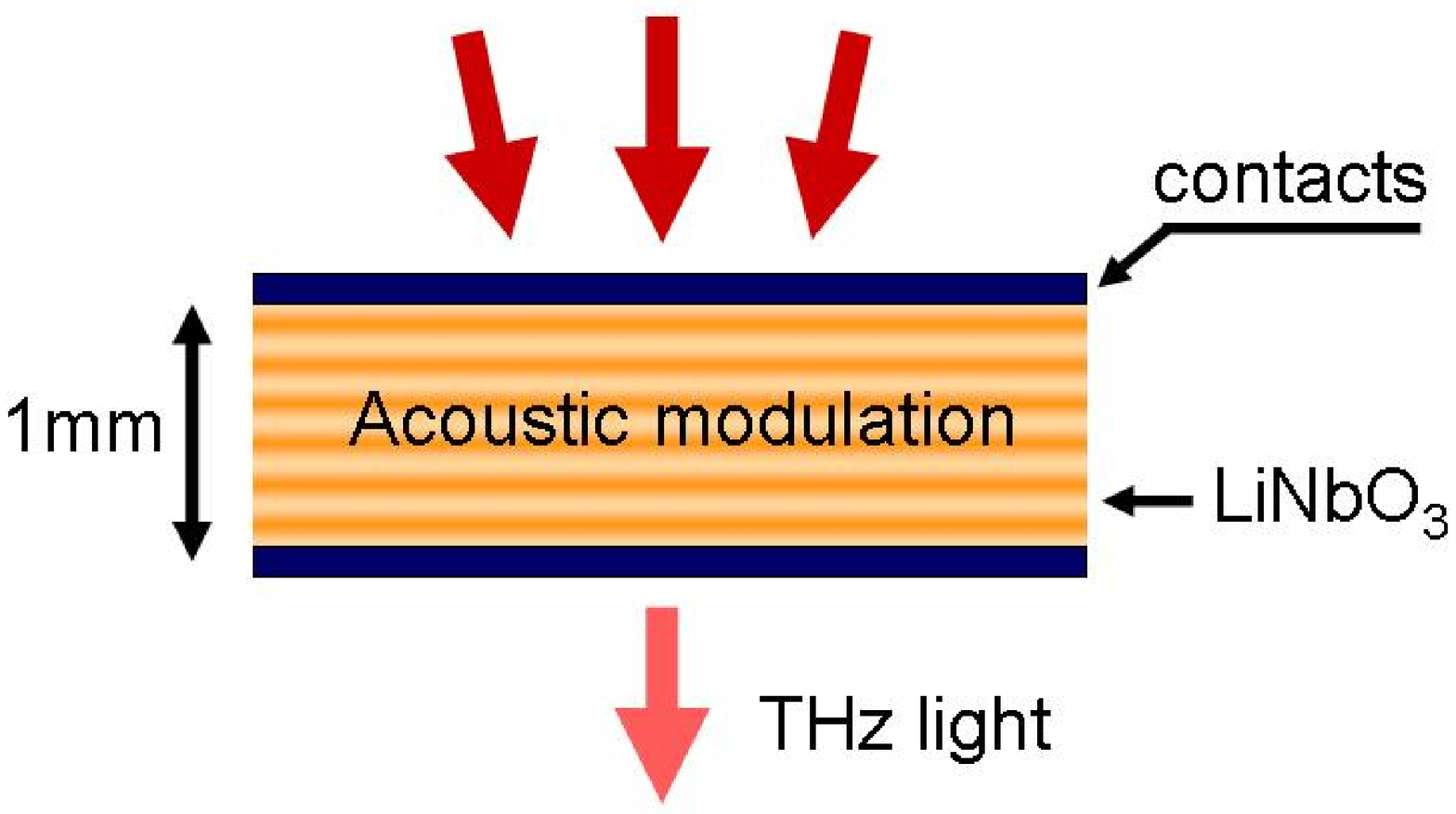}
\vskip-0.2cm
    \caption{
(Color online) Schematic design of a LiNbO$_3$-based THz filter
with an acoustic resonator.
 }\label{fig:scheme}
\end{figure}

In this Letter we show that the concept of the resonant
acousto-optic (RAO) effect could be used to produce a tunable THz
filter. We present a design and model numerically a
LiNbO$_{3}$-based filter which is tunable in a wide range of up to
4\,THz. The RAO effect which we use is in general a mediation of
the interaction between an acoustic wave and a light field by a
solid state excitations, when the light field and the excitation
are in
resonance.\cite{Ivanov2001,Ivanov2005,Ivanov2008,Poolman2010,Muljarov2010}
In ionic crystals with soft TO-phonon modes, the interaction
between a coherent acoustic wave and the THz field is mediated by
a TO phonon.\cite{Poolman2010,Muljarov2010} The TO phonon is
coupled to the acoustic wave via the anharmonicity present in the
interatomic potential of the crystal lattice. This makes
LiNbO$_{3}$ an excellent candidate for the filter, as it is both
ionic and very strongly anharmonic. In fact, the strength of the
anharmonicity in LiNbO$_{3}$ is such that both the cubic and
quartic terms of the Taylor expansion of the interatomic potential
have to be considered.

The design of the LiNbO$_{3}$-based tunable filter is shown
schematically in Fig.\,1. A thin slab of LiNbO$_{3}$ (1\,mm thick
or less) is placed between doped semiconductor contacts producing
a resonator for ultrasound waves but at the same time transparent
for THz light. Owing to the phonon anharmonicities, two
counter-propagating TO-phonon polariton waves in LiNbO$_{3}$
excited by an external THz light can strongly couple to each other
through the coherent acoustic field in the resonator.  This
happens at specific frequencies determined by the resonant Bragg
condition.\cite{Muljarov2010} At these resonant frequencies,
acoustically induced transitions between TO-phonon states result
in band gaps in the polariton dispersion creating peaks in the
reflectivity. The anharmonic scattering channels responsible for
these transitions are ``TO~phonon $\pm$ one
(two)~longitudional~acoustic~phonon(s)$ \rightarrow$ TO~phonon'',
for the cubic (quartic) anharmonicity. The efficiency of these
channels determines the width of the acoustically induced band
gaps and the reflectivity spikes and increases with increasing
acoustic intensity $I_{\rm ac}$.\cite{Poolman2010}

To model numerically the RAO effect in the LiNbO$_{3}$ filter we
calculate reflectivity from the acoustically modulated
semiconductor as well as extinction of the THz light inside a
LiNbO$_{3}$ crystal. The acoustic wave parametrically modulates
the semiconductor through the macroscopic polarization ${\bf
P}({\bf r},t)$ of TO phonons. This modulation can be expressed
mathematically using the Hopfield model\cite{Hopfield1968} of
polaritons, generalized to the case of a parametric perturbation
by an external coherent acoustic wave. The material equation for
the TO-phonon polarization is then extended to the
form\cite{Poolman2010}
\begin{align}
&\frac{\varepsilon_{\rm b}}{4\pi}\omega_{R}^{2}{\bf E}({\bf r},t)=
\left[\spderiv{}{t} + 2\gamma_{TO}\pderiv{}{ t} + \omega_{\rm
TO}^{2}\right] {\bf P}({\bf r},t)\nonumber \\
    & +2\omega_{\rm TO}\nonumber\Big[ m_{3}I_{\rm ac}^{1/2}({\bf r},t) \cos(\Omega_{\rm ac}t - {\bf Kr + \phi_3}) +2m_{4}I_{\rm ac}({\bf r},t)\nonumber  \\
    &   + m_{4}I_{\rm ac}({\bf r},t) \cos(2\Omega_{\rm ac}t - 2{\bf Kr}+2\phi_4)
        \Big] {\bf P}({\bf r},t)
    \label{eq:MacroP}
\end{align}
which includes both cubic and quartic terms of the coupling to the
ultrasound wave, and in combination with Maxwell's equation and
boundary conditions allows us to calculate the macroscopic
electric field ${\bf E}({\bf r},t)$. The standard material
parameters $\omega_{\rm TO}$, $\gamma_{\rm TO}$, and $\omega_{\rm
R}$ are the TO-phonon frequency, room temperature phenomenological
TO-phonon damping, and the Rabi frequency of the Hopfield
polariton, respectively. The background dielectric constant is
given by $\varepsilon_{b}$. There are two controlling parameters
for the RAO effect, the frequency of the acoustic wave
$\Omega_{\rm ac}=\nu_{\rm ac}/(2\pi)=v_s K$, where $K=|{\bf K}|$
is the acoustic wave vector, and the strength of the coupling
between the acoustic wave and the TO-phonon field. This coupling
strength is controlled by the acoustic intensity $I_{\rm ac}({\bf
r},t)$ and is given by $m_{3}I_{\rm ac}^{1/2}$ for cubic and
$m_{4}I_{\rm ac}$ for quartic anharmonicity, the amplitudes of
intrinsic coupling constants being $m_{3}=1.52\times 10^7\, {\rm
s^{1/2}g^{-1/2}}$ and $m_{4}=1.64\,{\rm s^{2}g^{-1}}$ in
LiNbO$_{3}$. The coupling constants are derived from the potential
of the anharmonic components added to the Hopfield
Hamiltonian.\cite{Poolman2010,Romero-Rochin1999,Abrahams2002,Menendez1984}
In this paper, the coupling to a plane bulk acoustic wave with a
moderate intensity $I_{\rm ac}({\bf r},t)=I_{\rm ac}=25 {\rm
kW/cm^{2}}$ is considered.  The values of the phase shift
$\phi=\phi_3-\phi_4$ between the complex anharmonic coupling
parameters in Eq.~\ref{eq:MacroP} are not available in the
literature. Despite the large differences in coupling strength
($\hbar m_{3}I_{\rm ac}^{1/2}=5.02\,$\,meV while $\hbar
m_{4}I_{\rm ac}= 0.27\,$\,meV) and the fact that the quartic
modulation alone leads to negligibly small changes in the
LiNbO$_{3}$ reflection, the cubic and quartic modulations
interfere with each other resulting in noticeable $\phi$-dependent
effects which are also discussed below.

The modulation of the medium due to either cubic or quartic
anharmonicities has been studied by us in
Refs.\,\onlinecite{Poolman2010, Muljarov2010}. In this Letter, we
take into account both cubic and quartic modulations on an equal
footing, including their interference. For collinear propagation
of a THz light and an ultrasound wave, we calculate the Bragg
scattering of the light by the acoustic wave, which is resonantly
enhanced by the TO-phonon component of the polariton in
LiNbO$_{3}$. To do so, we use the numerical method developed in
Refs.\,\onlinecite{Poolman2010}. In particular, we solve
numerically Eq.~\ref{eq:MacroP} in a semi-infinite LiNbO$_{3}$
region, using a superposition of partial polaritonic plane waves
with all important Bragg harmonics taken into account. Because of
the large strength of the mediated coupling we take into up to 400
harmonics. Applying Maxwellian boundary conditions we calculate
the amplitudes of the partial waves of the transmitted and
reflected electric field. Full details of the calculation method
can be found in Ref.\,\onlinecite{Muljarov2010}.

\begin{figure}[t]
    \includegraphics[scale=0.85]{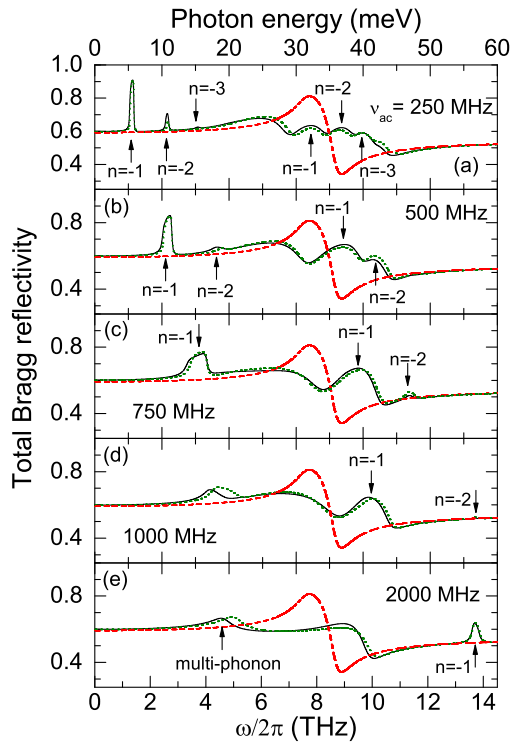}
    \caption{
(Color online) (a) to (e) show the development the THz reflection
of LiNbO$_{3}$ with varying acoustic frequency and different phase
shift between the cubic and quartic anharmonic parameters,
$\phi=0$ (black solid lines) and $\phi=\pi/2$ (green dated lines).
The red dashed line is the reflectivity spectrum at $I_{\rm
ac}=0$. In both cases $\omega_{\rm TO}=31$\,meV,
$\omega_{R}=18.0$\,meV, $\gamma_{\rm TO}=1.26$\,meV, and
$\varepsilon_b=44$.\cite{Abrahams2002}
 }
    \label{fig:freq}
\end{figure}

The total Bragg reflectivity which includes all important Bragg
replicas, calculated for the extreme cases of constructive
($\phi=0$) and destructive interference ($\phi=\pi/2$) is compared
in Fig.~\ref{fig:freq} with the TO-phonon polariton reflectivity
at $I_{\rm ac}=0$ (red dashed lines). The arrows indicate the
peaks in the reflectivity caused by the gaps in the quasi-energy
spectrum due to acoustically induced transitions between different
TO-phonon polaritons. The peaks occur at the resonant Bragg
condition $\omega=\omega_{\rm pol}(nK/2)$ [$\omega_{\rm pol}(k)$
is the TO-phonon polariton quasi-energy dispersion], both in the
lower polariton (LP) and in the upper polariton (UP) branches, at
the frequencies where co- and contra-propagating Bragg scattered
polariton states are brought into
resonance.\cite{Poolman2010,Muljarov2010}

The dramatic changes of the reflectivity peaks in a wide spectral
range and, as suggested by the Bragg condition, their blue shift
with increasing $\nu_{\rm ac}$ demonstrated in
Fig.~\ref{fig:freq}, is the phenomenon which we propose to use for
a widely tunable THz filter. The acoustic frequency provides an
easy to manipulate parameter which allows for the tuning of the
reflectivity. It can be clearly seen in Fig.~\ref{fig:freq}~(a-d)
that the light frequency range of $1\leq\omega/(2\pi)\leq4\,{\rm
THz}$ can be covered by the acoustic modulation of the LP branch
with up to $\nu_{\rm ac}\sim1\,{\rm GHz}$.

The $n$-th peak in the reflection physically corresponds to
stimulated emission of $|n|$ dressed acoustic phonons and to
down-conversion of the incoming THz light. In the LP branch, such
dressing with higher-order processes of re-emission and
re-absorption of acoustic phonons by the TO-phonon polariton
becomes more and more efficient with increasing acoustic frequency
$\nu_{\rm ac}$, due to the flattening of the polariton dispersion
near the bottom of the Restrahlen band and increasing of the
TO-phonon fraction in the polariton states, which enhances their
coupling to the acoustic wave. This results in widening of the
lower peak with $\nu_{\rm ac}$ and even more pronounced spectral
changes in the Restrahlen band itself, see Fig.~\ref{fig:freq},
(a-e).

\begin{figure}[t]
    \includegraphics[scale=0.85]{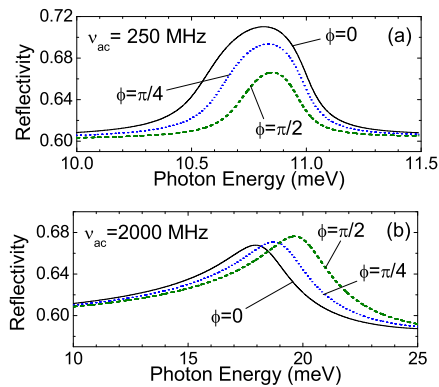}
    \caption{
(Color online) (a) The total Bragg reflectivity with the emphasis
on the $n=-2$ Bragg signal in LP branch calculated for $\nu_{\rm
ac}=250~{\rm MHz}$ and different phase: $\phi=0$, $\pi/4$, and
$\pi/2$. (b) The cumulative cups-like peak in the reflectivity
spectrum calculated for $\nu_{\rm ac}=2~{\rm GHz}$. }
    \label{fig:phase}
\end{figure}

\begin{figure}[t]
    \includegraphics[scale=0.8]{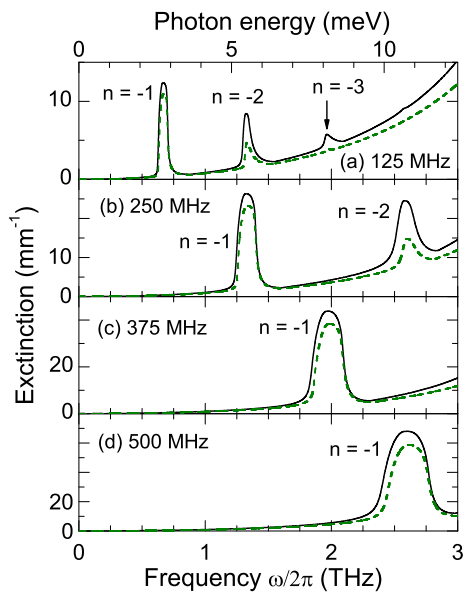}
    \caption{
(Color online) Extinction of the electric field at different
acoustic frequencies, for $\phi=0$ (black solid lines) and
$\phi=\pi/2$ (green dashed lines).}
    \label{fig:extinction}
\end{figure}

From Eq.~\ref{eq:MacroP} it is clear that the phase shift $\phi$
will only affect the peak which is caused by both cubic and
quartic anharmonicity. The cubic anharmonicity contributes to all
Bragg signals ($n=-1$, $n=-2$, and so on), the quartic
anharmonicity contributes to even $n$ only. This is why the
strongest interference is observed for $n=-2$ Bragg signal which
is detailed in Fig.\,\ref{fig:phase}~(a) for different values of
$\phi$. The peak position remains practically unaffected by $\phi$
as it is determined within this acoustic frequency range solely by
the resonant Bragg condition. For high acoustic frequency, shown
in Fig.~\ref{fig:phase}~(b), the line width and height of the LP
peak are not greatly effected by the phase. However, the peak
position is strongly $\phi$-dependent because it is no longer
determined by the Bragg condition. Instead, the individual
acoustically induced band gaps accumulate at the bottom edge of
the Restrahlen band to produce a cusp-like feature seen in
Fig.~\ref{fig:phase}~(b), which is discussed in detail in
Ref.\,\onlinecite{Muljarov2010}.

To quantify the transmission through the LiNbO$_{3}$ slab we plot
in Fig.\,4 the extinction which is calculated as an inverse decay
length of the electric field inside the slab. The spectra show a
high contrast in THz light filtering as well as its wide-range
tunability that demonstrates a large figure of merit of the
proposed device.

In conclusion, our numerical modelling of the RAO effect in
LiNbO$_{3}$ shows that there is potential for a THz filter that
can be continuously tuned between 0.5\,THz and $\sim4\,$THz. This
could be achieved through the manipulation of the spectral
position and height of peaks in reflectivity and extinction via
the acousto-optical resonant Bragg condition.

The authors thank S.\,G.\,Tikhodeev and W. Langbein for valuable
discussions. The work was supported by the Royal Society (Grant
JP0766306), EPSRC, and WIMCS.

\bibliography{APL}

\providecommand{\noopsort}[1]{}\providecommand{\singleletter}[1]{#1}%
\begin{thebibliography}{22}%
\makeatletter
\providecommand \@ifxundefined [1]{%
 \@ifx{#1\undefined}
}%
\providecommand \@ifnum [1]{%
 \ifnum #1\expandafter \@firstoftwo
 \else \expandafter \@secondoftwo
 \fi
}%
\providecommand \@ifx [1]{%
 \ifx #1\expandafter \@firstoftwo
 \else \expandafter \@secondoftwo
 \fi
}%
\providecommand \natexlab [1]{#1}%
\providecommand \enquote  [1]{``#1''}%
\providecommand \bibnamefont  [1]{#1}%
\providecommand \bibfnamefont [1]{#1}%
\providecommand \citenamefont [1]{#1}%
\providecommand \href@noop [0]{\@secondoftwo}%
\providecommand \href [0]{\begingroup \@sanitize@url \@href}%
\providecommand \@href[1]{\@@startlink{#1}\@@href}%
\providecommand \@@href[1]{\endgroup#1\@@endlink}%
\providecommand \@sanitize@url [0]{\catcode `\\12\catcode `\$12\catcode
  `\&12\catcode `\#12\catcode `\^12\catcode `\_12\catcode `\%12\relax}%
\providecommand \@@startlink[1]{}%
\providecommand \@@endlink[0]{}%
\providecommand \url  [0]{\begingroup\@sanitize@url \@url }%
\providecommand \@url [1]{\endgroup\@href {#1}{\urlprefix }}%
\providecommand \urlprefix  [0]{URL }%
\providecommand \Eprint [0]{\href }%
\providecommand \doibase [0]{http://dx.doi.org/}%
\providecommand \selectlanguage [0]{\@gobble}%
\providecommand \bibinfo  [0]{\@secondoftwo}%
\providecommand \bibfield  [0]{\@secondoftwo}%
\providecommand \translation [1]{[#1]}%
\providecommand \BibitemOpen [0]{}%
\providecommand \bibitemStop [0]{}%
\providecommand \bibitemNoStop [0]{.\EOS\space}%
\providecommand \EOS [0]{\spacefactor3000\relax}%
\providecommand \BibitemShut  [1]{\csname bibitem#1\endcsname}%
\let\auto@bib@innerbib\@empty
\bibitem [{\citenamefont {Schmuttenmaer}(2004)}]{Schmuttenmaer2004}%
  \BibitemOpen
  \bibfield  {author} {\bibinfo {author} {\bibfnamefont {C.~A.}\ \bibnamefont
  {Schmuttenmaer}},\ }\href@noop {} {\bibfield  {journal} {\bibinfo  {journal}
  {Chem.\ Rev.}\ }\textbf {\bibinfo {volume} {104}},\ \bibinfo {pages} {1759}
  (\bibinfo {year} {2004})}\BibitemShut {NoStop}%
\bibitem [{\citenamefont {Feurer}, \citenamefont {Vaughan},\ and\ \citenamefont
  {Nelson}(2003)}]{Feurer2003}%
  \BibitemOpen
  \bibfield  {author} {\bibinfo {author} {\bibfnamefont {J.~C.}\ \bibnamefont
  {Feurer}}, \bibinfo {author} {\bibfnamefont {J.~C.}\ \bibnamefont {Vaughan}},
  \ and\ \bibinfo {author} {\bibfnamefont {K.~A.}\ \bibnamefont {Nelson}},\
  }\href@noop {} {\bibfield  {journal} {\bibinfo  {journal} {Science}\ }\textbf
  {\bibinfo {volume} {299}},\ \bibinfo {pages} {374} (\bibinfo {year}
  {2003})}\BibitemShut {NoStop}%
\bibitem [{\citenamefont {Kojima}\ \emph {et~al.}(2003)\citenamefont {Kojima},
  \citenamefont {Tsumura}, \citenamefont {Takeda},\ and\ \citenamefont
  {Nishizawa}}]{Kojima2003}%
  \BibitemOpen
  \bibfield  {author} {\bibinfo {author} {\bibfnamefont {S.}~\bibnamefont
  {Kojima}}, \bibinfo {author} {\bibfnamefont {N.}~\bibnamefont {Tsumura}},
  \bibinfo {author} {\bibfnamefont {W.~M.}\ \bibnamefont {Takeda}}, \ and\
  \bibinfo {author} {\bibfnamefont {S.}~\bibnamefont {Nishizawa}},\ }\href@noop
  {} {\bibfield  {journal} {\bibinfo  {journal} {Phys. Rev. B}\ }\textbf
  {\bibinfo {volume} {67}},\ \bibinfo {pages} {035102} (\bibinfo {year}
  {2003})}\BibitemShut {NoStop}%
\bibitem [{\citenamefont {Han}\ and\ \citenamefont {Zhang}(2001)}]{Han2001}%
  \BibitemOpen
  \bibfield  {author} {\bibinfo {author} {\bibfnamefont {P.~Y.}\ \bibnamefont
  {Han}}\ and\ \bibinfo {author} {\bibfnamefont {X.-C.}\ \bibnamefont
  {Zhang}},\ }\href@noop {} {\bibfield  {journal} {\bibinfo  {journal} {Meas.
  Sci. amd Technol.}\ }\textbf {\bibinfo {volume} {12}},\ \bibinfo {pages}
  {1747} (\bibinfo {year} {2001})}\BibitemShut {NoStop}%
\bibitem [{\citenamefont {Stoyanov}\ \emph {et~al.}(2002)\citenamefont
  {Stoyanov}, \citenamefont {Ward}, \citenamefont {Feurer},\ and\ \citenamefont
  {Nelson}}]{Stoyanov2002}%
  \BibitemOpen
  \bibfield  {author} {\bibinfo {author} {\bibfnamefont {N.~S.}\ \bibnamefont
  {Stoyanov}}, \bibinfo {author} {\bibfnamefont {D.~W.}\ \bibnamefont {Ward}},
  \bibinfo {author} {\bibfnamefont {T.}~\bibnamefont {Feurer}}, \ and\ \bibinfo
  {author} {\bibfnamefont {K.~A.}\ \bibnamefont {Nelson}},\ }\href@noop {}
  {\bibfield  {journal} {\bibinfo  {journal} {Nature Mat.}\ }\textbf {\bibinfo
  {volume} {1}},\ \bibinfo {pages} {95} (\bibinfo {year} {2002})}\BibitemShut
  {NoStop}%
\bibitem [{\citenamefont {Siegel}(2002)}]{Siegel2002}%
  \BibitemOpen
  \bibfield  {author} {\bibinfo {author} {\bibfnamefont {P.~H.}\ \bibnamefont
  {Siegel}},\ }\href@noop {} {\bibfield  {journal} {\bibinfo  {journal} {IEEE
  T. Microw. Theory}\ }\textbf {\bibinfo {volume} {50}},\ \bibinfo {pages}
  {910} (\bibinfo {year} {2002})}\BibitemShut {NoStop}%
\bibitem [{\citenamefont {Fischer}\ \emph {et~al.}(2005)\citenamefont
  {Fischer}, \citenamefont {Hoffmann}, \citenamefont {Helm}, \citenamefont
  {Modjesch},\ and\ \citenamefont {Jepsen}}]{Fischer2005}%
  \BibitemOpen
  \bibfield  {author} {\bibinfo {author} {\bibfnamefont {B.}~\bibnamefont
  {Fischer}}, \bibinfo {author} {\bibfnamefont {M.}~\bibnamefont {Hoffmann}},
  \bibinfo {author} {\bibfnamefont {H.}~\bibnamefont {Helm}}, \bibinfo {author}
  {\bibfnamefont {G.}~\bibnamefont {Modjesch}}, \ and\ \bibinfo {author}
  {\bibfnamefont {P.~U.}\ \bibnamefont {Jepsen}},\ }\href@noop {} {\bibfield
  {journal} {\bibinfo  {journal} {Semicond. Sci. Technol.}\ }\textbf {\bibinfo
  {volume} {20}},\ \bibinfo {pages} {S246} (\bibinfo {year}
  {2005})}\BibitemShut {NoStop}%
\bibitem [{\citenamefont {Federici}\ \emph {et~al.}(2005)\citenamefont
  {Federici}, \citenamefont {Brian~Schulkin}, \citenamefont {Huang},
  \citenamefont {Gary}, \citenamefont {Barat}, \citenamefont {Oliveira},\ and\
  \citenamefont {Zimdars}}]{Federici2005}%
  \BibitemOpen
  \bibfield  {author} {\bibinfo {author} {\bibfnamefont {J.~F.}\ \bibnamefont
  {Federici}}, \bibinfo {author} {\bibfnamefont {B.}~\bibnamefont
  {Brian~Schulkin}}, \bibinfo {author} {\bibfnamefont {F.}~\bibnamefont
  {Huang}}, \bibinfo {author} {\bibfnamefont {D.}~\bibnamefont {Gary}},
  \bibinfo {author} {\bibfnamefont {R.}~\bibnamefont {Barat}}, \bibinfo
  {author} {\bibfnamefont {F.}~\bibnamefont {Oliveira}}, \ and\ \bibinfo
  {author} {\bibfnamefont {D.}~\bibnamefont {Zimdars}},\ }\href@noop {}
  {\bibfield  {journal} {\bibinfo  {journal} {Semicond. Sci. Technol.}\
  }\textbf {\bibinfo {volume} {20}},\ \bibinfo {pages} {S266} (\bibinfo {year}
  {2005})}\BibitemShut {NoStop}%
\bibitem [{\citenamefont {Libon}\ \emph {et~al.}(2008)\citenamefont {Libon},
  \citenamefont {Baumg$\rm\ddot{a}$rtner}, \citenamefont {Hemple},
  \citenamefont {Hecker}, \citenamefont {Feldmann}, \citenamefont {Kock},\ and\
  \citenamefont {Dawson}}]{Libon2000}%
  \BibitemOpen
  \bibfield  {author} {\bibinfo {author} {\bibfnamefont {I.~H.}\ \bibnamefont
  {Libon}}, \bibinfo {author} {\bibfnamefont {S.}~\bibnamefont
  {Baumg$\rm\ddot{a}$rtner}}, \bibinfo {author} {\bibfnamefont
  {M.}~\bibnamefont {Hemple}}, \bibinfo {author} {\bibfnamefont {N.~E.}\
  \bibnamefont {Hecker}}, \bibinfo {author} {\bibfnamefont {J.}~\bibnamefont
  {Feldmann}}, \bibinfo {author} {\bibfnamefont {M.}~\bibnamefont {Kock}}, \
  and\ \bibinfo {author} {\bibfnamefont {P.}~\bibnamefont {Dawson}},\
  }\href@noop {} {\bibfield  {journal} {\bibinfo  {journal} {Opt. Lett.}\
  }\textbf {\bibinfo {volume} {33}},\ \bibinfo {pages} {1401} (\bibinfo {year}
  {2008})}\BibitemShut {NoStop}%
\bibitem [{\citenamefont {Pan}\ \emph {et~al.}(2005)\citenamefont {Pan},
  \citenamefont {Hsieh}, \citenamefont {Pan}, \citenamefont {Tanaka},
  \citenamefont {Miyamaru}, \citenamefont {Tani},\ and\ \citenamefont
  {Hangyo}}]{Pan2005}%
  \BibitemOpen
  \bibfield  {author} {\bibinfo {author} {\bibfnamefont {C.-L.}\ \bibnamefont
  {Pan}}, \bibinfo {author} {\bibfnamefont {C.-F.}\ \bibnamefont {Hsieh}},
  \bibinfo {author} {\bibfnamefont {R.-P.}\ \bibnamefont {Pan}}, \bibinfo
  {author} {\bibfnamefont {M.}~\bibnamefont {Tanaka}}, \bibinfo {author}
  {\bibfnamefont {F.}~\bibnamefont {Miyamaru}}, \bibinfo {author}
  {\bibfnamefont {M.}~\bibnamefont {Tani}}, \ and\ \bibinfo {author}
  {\bibfnamefont {M.}~\bibnamefont {Hangyo}},\ }\href@noop {} {\bibfield
  {journal} {\bibinfo  {journal} {Opt. Express.}\ }\textbf {\bibinfo {volume}
  {13}},\ \bibinfo {pages} {113921} (\bibinfo {year} {2005})}\BibitemShut
  {NoStop}%
\bibitem [{\citenamefont {Chen}\ \emph {et~al.}(2006)\citenamefont {Chen},
  \citenamefont {Pan}, \citenamefont {Hsieh}, \citenamefont {Lin},\ and\
  \citenamefont {Pan}}]{Chen2006}%
  \BibitemOpen
  \bibfield  {author} {\bibinfo {author} {\bibfnamefont {C.-Y.}\ \bibnamefont
  {Chen}}, \bibinfo {author} {\bibfnamefont {C.-L.}\ \bibnamefont {Pan}},
  \bibinfo {author} {\bibfnamefont {C.-F.}\ \bibnamefont {Hsieh}}, \bibinfo
  {author} {\bibfnamefont {Y.-F.}\ \bibnamefont {Lin}}, \ and\ \bibinfo
  {author} {\bibfnamefont {R.-P.}\ \bibnamefont {Pan}},\ }\href@noop {}
  {\bibfield  {journal} {\bibinfo  {journal} {Appl. Phys. Lett.}\ }\textbf
  {\bibinfo {volume} {88}},\ \bibinfo {pages} {101107} (\bibinfo {year}
  {2006})}\BibitemShut {NoStop}%
\bibitem [{\citenamefont {Ho}\ \emph {et~al.}(2008)\citenamefont {Ho},
  \citenamefont {Pan}, \citenamefont {Hsieh},\ and\ \citenamefont
  {Pan}}]{Ho2008}%
  \BibitemOpen
  \bibfield  {author} {\bibinfo {author} {\bibfnamefont {I.-C.}\ \bibnamefont
  {Ho}}, \bibinfo {author} {\bibfnamefont {C.-L.}\ \bibnamefont {Pan}},
  \bibinfo {author} {\bibfnamefont {C.-F.}\ \bibnamefont {Hsieh}}, \ and\
  \bibinfo {author} {\bibfnamefont {R.-P.}\ \bibnamefont {Pan}},\ }\href@noop
  {} {\bibfield  {journal} {\bibinfo  {journal} {Opt. Lett.}\ }\textbf
  {\bibinfo {volume} {33}},\ \bibinfo {pages} {1401} (\bibinfo {year}
  {2008})}\BibitemShut {NoStop}%
\bibitem [{\citenamefont {Drysdale}\ \emph {et~al.}(2004)\citenamefont
  {Drysdale}, \citenamefont {Gregory}, \citenamefont {Baker}, \citenamefont
  {Linfield}, \citenamefont {Tribe},\ and\ \citenamefont
  {Cumming}}]{Drysdale2004}%
  \BibitemOpen
  \bibfield  {author} {\bibinfo {author} {\bibfnamefont {T.~D.}\ \bibnamefont
  {Drysdale}}, \bibinfo {author} {\bibfnamefont {I.~S.}\ \bibnamefont
  {Gregory}}, \bibinfo {author} {\bibfnamefont {C.}~\bibnamefont {Baker}},
  \bibinfo {author} {\bibfnamefont {E.~H.}\ \bibnamefont {Linfield}}, \bibinfo
  {author} {\bibfnamefont {W.~R.}\ \bibnamefont {Tribe}}, \ and\ \bibinfo
  {author} {\bibfnamefont {D.~R.~S.}\ \bibnamefont {Cumming}},\ }\href@noop {}
  {\bibfield  {journal} {\bibinfo  {journal} {Appl. Phys. Lett.}\ }\textbf
  {\bibinfo {volume} {85}},\ \bibinfo {pages} {5173} (\bibinfo {year}
  {2004})}\BibitemShut {NoStop}%
\bibitem [{\citenamefont {Ivanov}\ and\ \citenamefont
  {Littlewood}(2001)}]{Ivanov2001}%
  \BibitemOpen
  \bibfield  {author} {\bibinfo {author} {\bibfnamefont {A.~L.}\ \bibnamefont
  {Ivanov}}\ and\ \bibinfo {author} {\bibfnamefont {P.}~\bibnamefont
  {Littlewood}},\ }\href@noop {} {\bibfield  {journal} {\bibinfo  {journal}
  {Phys. Rev. Lett.}\ }\textbf {\bibinfo {volume} {87}},\ \bibinfo {pages}
  {136403} (\bibinfo {year} {2001})}\BibitemShut {NoStop}%
\bibitem [{\citenamefont {Ivanov}(2005)}]{Ivanov2005}%
  \BibitemOpen
  \bibfield  {author} {\bibinfo {author} {\bibfnamefont {A.~L.}\ \bibnamefont
  {Ivanov}},\ }\href@noop {} {\bibfield  {journal} {\bibinfo  {journal} {Phys.
  Status. Solidi A}\ }\textbf {\bibinfo {volume} {202}},\ \bibinfo {pages}
  {2657} (\bibinfo {year} {2005})}\BibitemShut {NoStop}%
\bibitem [{\citenamefont {Ivanov}(2008)}]{Ivanov2008}%
  \BibitemOpen
  \bibfield  {author} {\bibinfo {author} {\bibfnamefont {A.~L.}\ \bibnamefont
  {Ivanov}},\ }in\ \href@noop {} {\emph {\bibinfo {booktitle} {Problems of
  Condensed Matter Physics}}},\ \bibinfo {editor} {edited by\ \bibinfo {editor}
  {\bibfnamefont {A.~L.}\ \bibnamefont {Ivanov}}\ and\ \bibinfo {editor}
  {\bibfnamefont {S.~G.}\ \bibnamefont {Tikhodeev}}}\ (\bibinfo  {publisher}
  {Oxford University Press},\ \bibinfo {address} {Oxford},\ \bibinfo {year}
  {2008})\ pp.\ \bibinfo {pages} {301--322}\BibitemShut {NoStop}%
\bibitem [{\citenamefont {Poolman}, \citenamefont {Muljarov},\ and\
  \citenamefont {Ivanov}(2010)}]{Poolman2010}%
  \BibitemOpen
  \bibfield  {author} {\bibinfo {author} {\bibfnamefont {R.~H.}\ \bibnamefont
  {Poolman}}, \bibinfo {author} {\bibfnamefont {E.~A.}\ \bibnamefont
  {Muljarov}}, \ and\ \bibinfo {author} {\bibfnamefont {A.~L.}\ \bibnamefont
  {Ivanov}},\ }\href@noop {} {\bibfield  {journal} {\bibinfo  {journal} {Phys.
  Rev. B}\ }\textbf {\bibinfo {volume} {81}},\ \bibinfo {pages} {245208}
  (\bibinfo {year} {2010})}\BibitemShut {NoStop}%
\bibitem [{\citenamefont {Muljarov}, \citenamefont {Poolman},\ and\
  \citenamefont {Ivanov}(2011)}]{Muljarov2010}%
  \BibitemOpen
  \bibfield  {author} {\bibinfo {author} {\bibfnamefont {E.~A.}\ \bibnamefont
  {Muljarov}}, \bibinfo {author} {\bibfnamefont {R.~H.}\ \bibnamefont
  {Poolman}}, \ and\ \bibinfo {author} {\bibfnamefont {A.~L.}\ \bibnamefont
  {Ivanov}},\ }\href@noop {} {} (\bibinfo {year} {2011}),\ \bibinfo {note} {in
  print}\BibitemShut {NoStop}%
\bibitem [{\citenamefont {Hopfield}\ and\ \citenamefont
  {Thomas}(1963)}]{Hopfield1968}%
  \BibitemOpen
  \bibfield  {author} {\bibinfo {author} {\bibfnamefont {J.~J.}\ \bibnamefont
  {Hopfield}}\ and\ \bibinfo {author} {\bibfnamefont {D.~G.}\ \bibnamefont
  {Thomas}},\ }\href@noop {} {\bibfield  {journal} {\bibinfo  {journal} {Phys.
  Rev.}\ }\textbf {\bibinfo {volume} {132}},\ \bibinfo {pages} {563} (\bibinfo
  {year} {1963})}\BibitemShut {NoStop}%
\bibitem [{\citenamefont {Romero-Roch\'{\i}n}\ \emph
  {et~al.}(1999)\citenamefont {Romero-Roch\'{\i}n}, \citenamefont {Kochl},
  \citenamefont {Brennan},\ and\ \citenamefont {Nelson}}]{Romero-Rochin1999}%
  \BibitemOpen
  \bibfield  {author} {\bibinfo {author} {\bibfnamefont {V.}~\bibnamefont
  {Romero-Roch\'{\i}n}}, \bibinfo {author} {\bibfnamefont {R.~M.}\ \bibnamefont
  {Kochl}}, \bibinfo {author} {\bibfnamefont {C.~J.}\ \bibnamefont {Brennan}},
  \ and\ \bibinfo {author} {\bibfnamefont {K.~A.}\ \bibnamefont {Nelson}},\
  }\href@noop {} {\bibfield  {journal} {\bibinfo  {journal} {J. Chem. Phys.}\
  }\textbf {\bibinfo {volume} {111}},\ \bibinfo {pages} {3559} (\bibinfo {year}
  {1999})}\BibitemShut {NoStop}%
\bibitem [{\citenamefont {Abrahams}(2002)}]{Abrahams2002}%
  \BibitemOpen
  \bibfield  {author} {\bibinfo {author} {\bibfnamefont {S.~C.}\ \bibnamefont
  {Abrahams}},\ }in\ \href@noop {} {\emph {\bibinfo {booktitle} {Properties of
  Lithium Niobate}}},\ \bibinfo {editor} {edited by\ \bibinfo {editor}
  {\bibfnamefont {K.~K.}\ \bibnamefont {Wong}}}\ (\bibinfo  {publisher} {The
  Institute of Electrical Engineers},\ \bibinfo {year} {2002})\BibitemShut
  {NoStop}%
\bibitem [{\citenamefont {Men\'{e}ndez}\ and\ \citenamefont
  {Cardona}(1984)}]{Menendez1984}%
  \BibitemOpen
  \bibfield  {author} {\bibinfo {author} {\bibfnamefont {J.}~\bibnamefont
  {Men\'{e}ndez}}\ and\ \bibinfo {author} {\bibfnamefont {M.}~\bibnamefont
  {Cardona}},\ }\href@noop {} {\bibfield  {journal} {\bibinfo  {journal} {Phys.
  Rev. B}\ }\textbf {\bibinfo {volume} {29}},\ \bibinfo {pages} {2051}
  (\bibinfo {year} {1984})}\BibitemShut {NoStop}%
\end{thebibliography}%

\end{document}